\documentclass[aip,jcp,reprint,noshowkeys,superscriptaddress]{revtex4-2}
\usepackage{graphicx,dcolumn,bm,xcolor,microtype,multirow,amscd,amsmath,amssymb,amsfonts,physics,longtable,wrapfig,txfonts,siunitx,xspace}
\usepackage[version=4]{mhchem}

\usepackage[utf8]{inputenc}
\usepackage[T1]{fontenc}
\usepackage{txfonts}

\usepackage[
	colorlinks=true,
	citecolor=blue,
	breaklinks=true
	]{hyperref}
\usepackage{cancel}

\usepackage[normalem]{ulem}

\newcommand{\mc}{\multicolumn}

\newcommand{\HF}{\text{HF}}

\newcommand{\GW}{\text{$GW$}}
\newcommand{\GWC}{\text{$GW$+C}}
\newcommand{\C}{\text{C}}
\newcommand{\co}{\text{c}}
\newcommand{\QP}{\text{QP}}
\newcommand{\sat}{\text{sat}}

\newcommand{\hH}{\Hat{H}}

\newcommand{\hT}{\Hat{T}}

\newcommand{\bx}{\boldsymbol{x}}

\newcommand{\bO}{\boldsymbol{0}}
\newcommand{\bA}{\boldsymbol{A}}
\newcommand{\bB}{\boldsymbol{B}}

\newcommand{\bX}{\boldsymbol{X}}
\newcommand{\bY}{\boldsymbol{Y}}
\newcommand{\bOm}{\boldsymbol{\Omega}}

\newcommand{\eps}{\epsilon}

\newcommand{\Om}{\Omega}
\newcommand{\Sig}{\Sigma}
\newcommand{\sig}{\sigma}
\newcommand{\ii}{\mathrm{i}}
\newcommand{\SupMat}{\textcolor{blue}{supporting information}\xspace}

\newcommand{\LCPQ}{Laboratoire de Chimie et Physique Quantiques (UMR 5626), Universit\'e de Toulouse, CNRS, UPS, France}

\begin{document}	

\title{Cumulant Green's function methods for molecules}

\author{Pierre-Fran\c{c}ois \surname{Loos}}
	\email{loos@irsamc.ups-tlse.fr}
	\affiliation{\LCPQ}
\author{Antoine \surname{Marie}}
	\email{amarie@irsamc.ups-tlse.fr}
	\affiliation{\LCPQ}
\author{Abdallah \surname{Ammar}}
	\email{aammar@irsamc.ups-tlse.fr}
	\affiliation{\LCPQ}

\begin{abstract}
The cumulant expansion of the Green's function is a computationally efficient beyond-$GW$ approach renowned for its significant enhancement of satellite features in materials. 
In contrast to the ubiquitous $GW$ approximation of many-body perturbation theory, \textit{ab initio} cumulant expansions performed on top of $GW$ ($GW$+C) have demonstrated the capability to handle multi-particle processes by incorporating higher-order correlation effects or vertex corrections, yielding better agreements between experiment and theory for satellite structures.
While widely employed in condensed matter physics, very few applications of $GW$+C have been published on molecular systems.
Here, we assess the performance of this scheme on a series of 10-electron molecular systems (\ce{Ne}, \ce{HF}, \ce{H2O}, \ce{NH3}, and \ce{CH4}) where full configuration interaction estimates of the outer-valence quasiparticle and satellite energies are available.
\bigskip
\begin{center}
	\boxed{\includegraphics[width=0.5\linewidth]{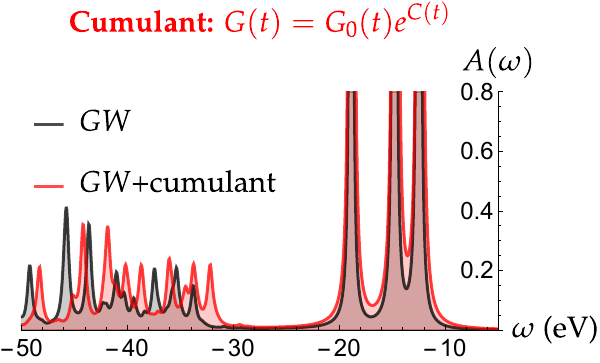}}
\end{center}
\end{abstract}

\maketitle

\section{Introduction}
\label{sec:intro}

The cumulant expansion is a versatile mathematical and theoretical tool that finds applications in various fields of physics. \cite{Kubo_1962} 
From a mathematical point of view, the cumulant expansion is an alternative to the moments for characterizing a probability distribution function.
It is often employed to obtain fluctuation and/or correlation functions beyond the mean-field approximation and is particularly valuable in dealing with many-body phenomena via the inclusion of higher-order effects.
It has been shown to be useful for understanding a wide range of physical phenomena across different energy scales, from the quantum realm to the cosmological scales. 
For example, cumulant-based approaches have been developed and applied in quantum field theory, \cite{Fauser_1996} statistical physics, \cite{Evans_1976} plasma physics, \cite{Cherif_2016} optics, \cite{Schack_1990} photonics, \cite{Sanchez-Barquilla_2020} cosmology, \cite{Garny_2023} and many others.

In condensed matter physics, the cumulant expansion has been particularly fruitful and is often used in the framework of Green's function theory. \cite{Hedin_1999,Onida_2002,MartinBook} Cumulant-based Green's function methods have been instrumental in providing a better description of satellite peaks in the context of photoemission spectroscopy of materials. \cite{Aryasetiawan_1996,Aryasetiawan_1998,Vos_1999,Vos_2001,Vos_2002,Kheifets_2003,Guzzo_2011,Lischner_2013,Gatt_2013,Guzzo_2014,Kas_2014,Caruso_2015b,Zhou_2015,Lischner_2015,Nakamur_2016,Gumhalter_2016,Kas_2016,Verdi_2017,Zhou_2018} 
Compared to the $GW$ approximation of many-body perturbation theory, \cite{Hedin_1965,Aryasetiawan_1998,Reining_2017,Golze_2019,Marie_2023b} the cumulant form yields better agreements between experimental observations and theoretical predictions for satellite structures, effectively reproducing the series of multiple satellites observed, for instance, in the photoemission spectrum of sodium \cite{Aryasetiawan_1996,Zhou_2015,Zhou_2018} and silicon. \cite{Kheifets_2003,Guzzo_2011,Lischner_2015,Caruso_2015a,Caruso_2015b,Gumhalter_2016,Vlcek_2018}
The cumulant expansion has also been succesfully employed to model x-ray photoemission spectra that probe core electrons. \cite{Nozieres_1969,Becrstedt_1980,Bechstedt_1980,Bechstedt_1982,Kas_2022a} This success can be understood thanks to the close connection of the cumulant ansatz with electron-boson models (see below). \cite{Lundqvist_1969,Langreth_1970,Gunnarsson_1994,Hedin_1980,Hedin_1999}
Indeed, satellites are many-body electronic excitations that go beyond the single-particle picture and the inclusion of higher-order correlation effects is required. \cite{Cederbaum_1974,Mejuto-Zaera_2021} 
In this context, cumulants are employed to approximate the higher-order terms (or vertex corrections) in the expansion of the one-body Green's function $G$.
Developments around \textit{ab initio} cumulant expansions are still ongoing and constitute an area of active research, \cite{Tzavala_2020,Cudazzo_2020a,Cudazzo_2020b,Kas_2022} especially for extensions to strongly correlated materials. \cite{Biermann_2016}

The fundamental concept behind the cumulant expansion is an exponential ansatz for the one-body Green's function in the time domain
\begin{equation} \label{eq:cum}
	G(t) = G_0(t) e^{C(t)}
\end{equation}
where $G_0(t)$ represents a reference Green's function and $C(t)$ is the so-called cumulant. This exponential ansatz shares obvious similarities with coupled-cluster (CC) theory \cite{ShavittBook} and quantum Monte Carlo. \cite{Foulkes_2001,Austin_2012} The cumulant expansion generates a Poisson series of satellites in the spectral function $A(\omega) = \pi^{-1} \abs{\Im G(\omega)}$, which establishes a direct link to experimental photoemission spectra via its connection to the photocurrent. \cite{Almbladh_1983,Hedin_1999,Guzzo_2011,Guzzo_2014,Zhou_2015,Zhou_2018,Kas_2022a} 
In practice, the central component of the cumulant is the $GW$ self-energy, aligning the procedure and computational cost of a cumulant calculation with that of $GW$. 
The $GW$+cumulant approach (henceforth $GW$+C) can thus be seen as an economical post-treatment that goes beyond the $GW$ approximation.

The cumulant has been also employed in realistic molecular systems \cite{Vlcek_2018,Vila_2020,Vila_2021} 
but much less than in solids. 
Our goal here is to assess how this approach performs in the context of molecular systems, where highly-accurate reference data for outer-valence quasiparticle and satellite energies are available. \cite{Marie_2024}

Our manuscript is organized as follows. In Sec.~\ref{sec:review}, we discuss several key developments and applications of the cumulant expansion in condensed matter physics before delving into the definition of the Green's function in Sec.~\ref{sec:G} and the derivation of the cumulant expansion in Sec.~\ref{sec:cumulant}. Section \ref{sec:GWC} reports a detailed derivation of the cumulant expansion based on the $GW$ self-energy. In Sec.~\ref{sec:res}, we compute quasiparticle and satellite energies on a series of 10-electron molecular systems (\ce{Ne}, \ce{HF}, \ce{H2O}, \ce{NH3}, and \ce{CH4}), as well as their corresponding spectral functions. Various comparisons are proposed to gauge the quality of these physical quantities. Our conclusions are drawn in Sec.~\ref{sec:ccl}

\section{Short Review}
\label{sec:review}

Fundamentally, the cumulant approach is rooted in electron-boson (or polaron) models where one or more fermions are coupled to bosons. In 1970, Langreth, \cite{Langreth_1970} building upon the work of Nozieres and De Dominicis, \cite{Nozieres_1969} studied a simple electron-boson model where a deep core electron is coupled with bosonic excitations. He successfully provided the exact solution to this model, the spectral function revealing a quasiparticle peak accompanied by a series of satellites following a Poissonian distribution, the corresponding Green's function having the form of Eq.~\eqref{eq:cum}. \cite{Langreth_1970}
This forms the basis of the cumulant ansatz.

One year before this, Lundqvist had shown that the approximate solution of this model, correct up to second order, is closely linked to the $GW$ approximation. \cite{Lundqvist_1969} The exact solution clearly evidences that the $GW$ approximation provides a good description of the quasiparticle peak but a poor description of the satellite region, $GW$ predicting one broad, wrongly-placed peak for the incoherent part of the spectrum. This so-called plasmaron, initially thought to be a novel type of quasiparticle excitation resulting from strong coupling between electrons and plasmons, \cite{Lundqvist_1969} was later attributed to a spurious solution of the Dyson equation or an artifact introduced by $GW$ which disappears at higher levels of theory. 

The plasmaron was actually first observed in the uniform electron gas (UEG) by Hedin, Lundqvist, and coworkers, \cite{Hedin_1967,Lundqvist_1967a,Lundqvist_1967b,Lundqvist_1968,Lundqvist_1969a} before being identified in the spectrum of core electrons. \cite{Lundqvist_1969} Despite claims and reports asserting the observation of the plasmaron, \cite{Bostwick_2010,Dial_2012} recent consensus suggests its non-existence in materials. The spurious prediction of the plasmaron highlights $GW$'s limitations in describing plasmon satellites. We refer the interested reader to Ref.~\onlinecite{Guzzo_PhD} for an exhaustive discussion about the plasmaron.

Langreth's polaron model provided crucial theoretical insights and found early applications in core-level spectroscopy. Building upon Langreth's work, Hedin \cite{Hedin_1980,Hedin_1991,Hedin_1999} went on to generalize the electron-boson model, occasionally termed a quasi-boson model, to the valence region. This extension includes dispersion or recoil effects, enhancing the model's applicability. Interestingly, Hedin's model shares close similarities with the electron-boson Hamiltonian used by T\"olle and Chan \cite{Tolle_2023} to highlight the connections between $GW$ and CC theory (see also Refs.~\onlinecite{Lange_2018,Quintero_2022}). 

In 1996, Aryasetiawan \textit{et al.} \cite{Aryasetiawan_1996} made a convincing comparison between the experimental valence photoemission spectra of sodium and aluminum and the spectra simulated via the \textit{ab initio} cumulant expansion. The shortcomings of $GW$ became evident, generating only one plasmon satellite at a considerably larger binding energy. In contrast, the cumulant expansion remarkably enhanced the spectral function, notably revealing multiple plasmon satellite structures and improving the satellite positions to align closely with experimental observations. However, the improvement in intensities was comparatively modest. One reason behind this observation is the absence of considerations for extrinsic and interference effects, which predominantly impact intensities.  
This study likely marks the first exploration of first principles cumulant-based calculations that specifically address satellite features, moving beyond the paradigmatic UEG model. Despite the cumulant's exactness for core ionizations, \cite{Langreth_1970,Almbladh_1983} the study reveals its surprising effectiveness for valence electrons.

One year later, Holm and Aryasetiawan conducted a thorough study on the impact of self-consistency within the cumulant approach applied to the UEG. \cite{Holm_1997} Comparing the self-consistent (sc) versions of $GW$ and $GW$+C revealed that (i) sc$GW$ and sc$GW$+C yield similar quasiparticle energies with a small impact of the self-consistency, (ii) while the effects on satellite positions is marginal at the sc$GW$+C level, there are notable modifications in intensity, and (iii) the improvement brought by self-consistency is more significant at the $GW$ level, the satellite structure becoming more realistic. In short, despite the small impact on satellites, the study highlights the nuanced effects of self-consistency within the cumulant approach, especially concerning the enhancement of plasmon satellite structure.

Following a period of relative quiet, Lucia Reining's and John Rehr's groups revived the general interest in $GW$+C, applying this approach to semiconductors, specifically silicon. \cite{Guzzo_2011} The spectral function produced by the cumulant expansion revealed multiple satellites in the valence band photoemission spectrum that eluded accurate description by $GW$ alone. This study reports the estimation of extrinsic, intrinsic, and interference effects, showcasing impressive agreement between theory and experiment. Indeed, to compare theoretical calculations with experimental data, the intrinsic spectral function alone is insufficient. Extrinsic losses, stemming from the scattering of the outgoing electron on its way to the detector, and interference effects between extrinsic and intrinsic contributions have to be considered. Notably, Ref.~\onlinecite{Guzzo_2011} highlights the shortcomings of $GW$ that were attributed to the plasmaron (see above). They also suggested that $GW$ might be more effective in cases where a sharp plasmaron is not formed, such as in the context of local plasmon structures within strongly correlated materials.

In a subsequent investigation, detailed in Ref.~\onlinecite{Guzzo_2014}, the group extended their work to graphite, maintaining a similar approach to Ref.~\onlinecite{Guzzo_2011}. Their data revealed multiple satellite replicas of intrinsic origin, augmented by extrinsic losses. Note that, in this material possessing more than one significant plasmon, there was no manifestation of a plasmaron at the $GW$ level.
In lower-dimensional materials such a doped graphene \cite{Lischner_2013} and the two-dimensional version of the UEG, \cite{Lischner_2014} Steven Louie's group showed that $GW$ also predicts a plasmaron while $GW$+C demonstrates commendable accuracy in reproducing satellite features.

In 2014, Kas, Rehr, and Reining introduced a variation of the cumulant expansion employing the retarded Green's function instead of the time-ordered Green's function, with a first application on the UEG. \cite{Kas_2014} In the time-ordered ansatz of the cumulant, there is a decoupling between electron and hole branches, yielding ``asymmetric'' spectral functions. The separation of the electron and hole branches proves justified for core levels. However, this becomes a significant limitation near the Fermi level. Overcoming this limitation is challenging, as attempts to surpass the basic cumulant often result in issues, such as negative spectral functions. \cite{Gunnarsson_1994} 

Another interesting paper from Reining's and Rehr's groups was published one year later. \cite{Zhou_2015} It deals with dynamical effects from a general perspective, focusing on the generation of new structures, such as satellites, arising from Coulomb interactions resulting from the coupling of excitations. The study presents a unified derivation of $GW$ and $GW$+C based on the equation-of-motion (EOM) formalism and reports a specific examination of bulk sodium in both valence and core regions, emphasizing the crucial role of self-consistency, particularly in the core region. The study demonstrates good agreement with experimental results when the intrinsic spectral function is adjusted for extrinsic and interference effects (see above). Although suitable for electron-hole satellites, the authors also point out the limitations $GW$+C for hole-hole satellites, as observed in nickel. \cite{Springer_1998} Interestingly, they outlined how one can apply the cumulant expansion to the two-body Green's function. This extension of the cumulant approach was further investigated and developed by Cudazzo and Reining to describe phenomena like double plasmon satellites or exciton-exciton coupling within the Bethe-Salpeter formalism. \cite{Cudazzo_2020a,Cudazzo_2020b}

In 2016, McClain \textit{et al.} \cite{McClain_2016} presented a notable study comparing the spectral function obtained at the CC level on finite-size UEGs with results from $GW$ and $GW$+C. One of the interesting points of this paper is the comparison of $GW$+C with another state-of-the-art method, CC with single and double excitations (CCSD). For the 14-electron UEG at Wigner-Seitz radius $r_s = 4$ (which corresponds approximately to the density of the valence electron in metallic sodium), they found, based on additional density-matrix renormalization group \cite{Chan_2011,Baiardi_2020} (DMRG) calculations and the inclusion of the triple excitations at the CC level, that CCSD performs better than both $GW$ and $GW$+C. Large errors were imputed to the underlying Hartree-Fock (HF) starting point which is known to be grossly inaccurate for metallic systems. Much better results were obtained by relying on a local-density approximation (LDA) starting point. Because, by construction, $GW$+C produces a plasmon-replica satellite structure even for finite systems, several satellite peaks with incorrect energies and underestimated peak heights were found. The picture is quite different for the 114-electron system, where $GW$ produces a single satellite peak too high in energy (the infamous spurious plasmaron), while $GW$+C@LDA and CCSD spectra were found to be qualitatively similar. However, the CCSD spectral function has a stronger quasiparticle peak with a larger spectral width, and more fine structure overall than the $GW$+C spectral function.

In the same timeframe, several notable developments emerged: (i) Vigil-Fowler \textit{et al.} \cite{Vigil-Fowler_2016} explored the dispersion and line shape of plasmon satellites across various systems, employing the retarded $GW$+C method to investigate systems with variable dimensionality (the one-dimensional UEG, two-dimensional doped graphene, the three-dimensional UEG, and silicon); (ii) Reichman's group \cite{Mayers_2016} proposed to use the improper self-energy instead of the usual proper self-energy in the framework of the retarded $GW$+C method; and (iii) Caruso and Giustino conducted a comprehensive analysis of the spectral function of the UEG, employing both $GW$ and $GW$+C methodologies, with a specific emphasis on angle-resolved spectral functions. \cite{Caruso_2015a,Caruso_2015b,Caruso_2016a} 

As a follow-up of their 2015 paper, \cite{Zhou_2015} Zhou \textit{et al.} provided a few years later an insightful comparison between the time-ordered and retarded cumulant expansions, with a special focus on the dispersion and intensity of the plasmon satellites in bulk sodium and the UEG. \cite{Zhou_2018} Although both approaches are exact for deep core electrons, the investigation reveals that small yet significant changes accumulate due to variations in the ansatz details, causing a significant shift in satellite positions. This evidences the intricate nature of satellite calculations, which were shown to be much more sensitive to those details than the quasiparticle energies. Factors such as the linear response approximation, the level of theory employed for computing the dynamical screening $W$, and the diagonal approximation of the self-energy were identified as particularly influential. The validity of the linear response approximation was later investigated by Tzavala \textit{et al.} \cite{Tzavala_2020}, who demonstrated how to derive $GW$+C beyond linear response, with the Kadanoff-Baym functional serving as the starting point for this derivation.

In 2018, Vleck presented, using a stochastic $GW$ approach, \cite{Neuhauser_2014,Vlcek_2017} one of the very few papers applying $GW$+C to small molecular systems (\ce{NH3}, \ce{PH3}, and \ce{C2H2}), although his primary focus remains on bulk-like (i.e.~large) silicon nanocrystals. \cite{Vlcek_2018} The comparison of the $GW$+C spectral functions with experimental photoemission spectra, as well as SAC-CI calculations, \cite{Wasada_1989,Ishida_2002} shows qualitative agreement with a significant weight transfer from the quasiparticle peak to the satellite region.

Finally, in 2020, Kowalski, Peng, and coworkers initiated developments of a real-time EOM-CC approach for cumulants, \cite{Rehr_2020,Vila_2020,Vila_2021,Vila_2022a,Vila_2022b,Pathak_2023} with applications to core excitations in small molecular systems. It is worth mentioning that their approach goes beyond the usual linear response of the cumulant expansion and yields, thanks to the CC exponential ansatz, a nonperturbative expression for the cumulant. They found that the nonlinear terms significantly improved the results, yielding accurate core binding energies as well as a satisfactory treatment of the absolute positions of the satellites and the overall shape of their feature, especially when double excitations are included.

\section{One-body Green's function}
\label{sec:G}

We denote as $G^\text{T}$ and $G^\text{R}$ the time-ordered and retarded Green's functions, \cite{Zhou_2018} and their matrix elements in the spinorbital basis are respectively defined as
\begin{subequations}
\begin{align}
	G_{pq}^\text{T}(t) & = (-\ii) \mel*{\Psi_0^N}{ \hT[a_p(t) a_q^\dag(0)] }{\Psi_0^N}
	\\
	G_{pq}^\text{R}(t) & = (-\ii) \Theta(t) \mel*{\Psi_0^N}{ \{ a_p(t),a_q^\dag(0) \} }{\Psi_0^N}
\end{align}
\end{subequations}
where $\hT$ and $\{ \cdot,\cdot \}$  are the time-ordering and anticommutation operators, respectively,
$\Theta(t)$ is the Heaviside function, and $\ket*{\Psi_0^N}$ is the exact $N$-electron ground-state wave function.
In the following, the indices $i$, $j$, $k$, and $l$ are occupied (hole) spinorbitals; $a$, $b$, $c$, and $d$ are unoccupied (particle) spinorbitals; $p$, $q$, $r$, and $s$ indicate arbitrary spinorbitals; and $\nu$ labels single (de)excitations.
Here $a_p(t)$ and $a_p^\dag(t)$ are annihilation and creation operators in the Heisenberg picture, i.e.,
\begin{align}
	a_{p}(t) & = e^{\ii \hH t} a_{p} e^{-\ii \hH t}
	&
	a_{p}^\dag(t) & = e^{\ii \hH t} a_{p}^\dag e^{-\ii \hH t}
\end{align}

Introducing the greater and lesser components of the Green's function,
\begin{subequations}
\begin{align}
	G_{pq}^>(t) & = - \ii \mel*{\Psi_0^N}{a_p(t) a_q^\dag(0)}{\Psi_0^N}
	\\
	G_{pq}^<(t) & = + \ii \mel*{\Psi_0^N}{a_p^\dag(t) a_q(0)}{\Psi_0^N}
\end{align}
\end{subequations}
the time-ordered and retarded Green's functions are expressed as 
\begin{subequations}
\begin{align}
	G_{pq}^\text{T}(t) & = \Theta(+t) G_{pq}^>(t) + \Theta(-t) G_{pq}^<(t)
	\\
	G_{pq}^\text{R}(t) & = \Theta(+t) G_{pq}^>(t) - \Theta(+t) G_{pq}^<(t)
\end{align}
\end{subequations}

In frequency space, the corresponding elements of the spectral function, which is directly related to the photoemission spectrum of the system (see Sec.~\ref{sec:review}), are given by
\begin{equation} \label{eq:Apq}
\begin{split}
	A_{pq}(\omega) 
	& = \frac{\ii}{2\pi} \qty[ G_{pq}^>(\omega) - G_{pq}^<(\omega) ]
	\\
	& = \frac{1}{\pi} \abs{ \Im G_{pq}^\text{T}(\omega) }
	= - \frac{1}{\pi} \Im G_{pq}^\text{R}(\omega) 
\end{split}
\end{equation}
The greater and lesser components of $G$ can be combined in various ways which are equivalent if observables are calculated consistently. \cite{Zhou_2018} 

In the time-ordered and retarded versions of the cumulant expansion, the diagonal matrix elements of $G^{\lessgtr}(t)$ share the same form
\begin{equation}
	G_{pp}^{\lessgtr}(t)  \propto e^{-\ii \eps_{p} t} e^{C_{pp}^{\lessgtr}(t)}
\end{equation}
where the $\eps_p$'s are the one-body energies employed to build $G_0(t)$ [see Eq.~\eqref{eq:cum}] and $C_{pp}^{\lessgtr}(t)$ are the cumulant matrix elements to be determined.
However, in the time-ordered formulation, because only hole (particle) states contribute to the electron removal (addition) spectrum, we have $G_{ii}^>(t) = 0$ [$G_{aa}^<(t) = 0$], which yields
\begin{subequations}
\begin{align}
	\label{eq:GiT}
	G_{ii}^\text{T}(t)  & = \ii \Theta(-t) e^{-\ii \eps_{i} t} e^{C_{ii}^\text{T}(t)}
	\\ 
	\label{eq:GaT}
	G_{aa}^\text{T}(t)  & = \ii \Theta(+t) e^{-\ii \eps_{a} t} e^{C_{aa}^\text{T}(t)}
\end{align}
\end{subequations}
For the retarded Green's function, hole and particle states contribute to both the electron addition and removal spectra, which explains the more symmetric definition of the following retarded Green's function
\begin{equation}
	G_{pp}^\text{R}(t)  = - \ii \Theta(t) e^{-\ii \eps_{p} t} e^{C_{pp}^\text{R}(t)}
\end{equation}

As mentioned in Sec.~\ref{sec:review}, in the time-ordered formulation of the cumulant, there is a separation between the electron and hole branches, as readily seen in Eqs.~\eqref{eq:GiT} and \eqref{eq:GaT}, leading to spectral functions that exhibit asymmetry. While this electron-hole branch decoupling is appropriate for core levels, it poses a notable limitation when approaching the Fermi level. Therefore, in the following, our derivation is based on the retarded Green's function.
As explained in Chapter 5 of Ref.~\onlinecite{MartinBook}, the poles of retarded quantities must be shifted to the lower part of the complex plane by an infinitesimal amount due to the requirement of causality. This is a major difference from the more conventional time-ordered quantities where the poles are displaced in the upper or lower plane depending on their relative position with respect to the chemical potential.

\section{The cumulant expansion}
\label{sec:cumulant}

The derivation of the cumulant expansion can be approached through different methods: diagrammatic techniques, \cite{Nozieres_1969,Hedin_1980} the EOM formalism, \cite{Almbladh_1983} the Baym-Kadanoff equation, \cite{Guzzo_2011} or by imposing its form and identifying the cumulant with the Dyson equation linking the Green's function and the self-energy. \cite{Gunnarsson_1994,Aryasetiawan_1996} The latter stands out as the simplest and we shall follow this procedure in our derivation. 
Additional details are provided in the \SupMat. It is also worth mentioning that, because we deal with small molecular systems, we rely on a HF reference Green's function. 

By definition, the cumulant ansatz of the Green's function is
\begin{equation} \label{eq:G}
	G(t)  = G^\HF(t) e^{C(t)}
\end{equation}
where, at the HF level, we have
\begin{equation} \label{eq:Gpq_HF}
	G_{pq}^\HF(t) = - \ii \Theta(t) e^{-\ii \eps_p^\HF t} \delta_{pq}
\end{equation}
and the $\eps^\HF_p$'s are HF one-electron orbital energies.

We start from the Dyson equation that links the Green's function to the HF Green's function via the correlation part of the self-energy
\begin{multline} \label{eq:Dyson_t}
  G(\bx_1\bx_{1'};t) = G^\HF(\bx_1\bx_{1'};t) \\
  + \int \dd{(t_1t_2)} \dd{(\bx_2\bx_3)} G^\HF(\bx_1\bx_{2};t-t_1) \Sig^\co(\bx_2\bx_{3};t_1-t_2) G(\bx_3\bx_{1'};t_2)
\end{multline}
where we have taken into account the time-translation invariance of the Hamiltonian and $\bx$ is a composite variable gathering spin and space.

Expanding Eq.~\eqref{eq:G} as a function of $C(t)$ and Eq.~\eqref{eq:Dyson_t} with respect to $\Sig^\co$ yields
\begin{subequations}
\begin{align} 
	G(t) & = G^\HF(t) + G^\HF(t) C(t) + \cdots
	\\
	G(t) & = G^\HF(t) + \iint \dd{t_1} \dd{t_2} G^\HF(t-t_1) \Sig^\co(t_1-t_2) G^\HF(t_2)  + \cdots	
\end{align}
\end{subequations}
where we have omitted, for the sake of simplicity, the spin-spatial coordinates.
In order to determine $C(t)$ up to first order in $W$, the second terms of the right-hand side of the two previous equations are equated, which gives
\begin{equation}
	G^\HF(t) C(t) = \iint \dd{t_1} \dd{t_2} G^\HF(t-t_1) \Sig^\co(t_1-t_2) G^\HF(t_2)
\end{equation}
Projecting this expression in the spinorbital basis yields
\begin{equation}
	\sum_r G_{pr}^\HF(t) C_{rq}(t) = \sum_{rs} \iint \dd{t_1} \dd{t_2} G_{pr}^\HF(t-t_1) \Sig_{rs}^\co(t_1 - t_2) G_{sq}^\HF(t_2)
\end{equation}
or, by moving to frequency space,
\begin{equation}
	C_{pq}(t) = \ii \int \frac{\dd{\omega}}{2\pi} e^{-\ii(\omega-\eps_p^\HF)t} 
		G_{pp}^\HF(\omega) \Sig_{pq}^\co(\omega) G_{qq}^\HF(\omega)
\end{equation}
This last identity represents the general expression of the matrix elements of the cumulant within the linear response approximation. 

Using the expression of the elements of the HF retarded Green's function in frequency space

\begin{equation} \label{eq:GppHF}
	G_{pp}^\HF(\omega) = \lim_{\delta\to0^+}  \frac{1}{\omega - \qty(\eps_p^\HF - \ii \delta)}
\end{equation}
enforcing the diagonal approximation for the self-energy, i.e.~$\Sig_{pq}(\omega) \approx \delta_{pq} \Sig_{pp}(\omega)$, and applying a frequency shift, one gets
\begin{equation} \label{eq:Cpp}
	C_{pp}(t) = \ii \lim_{\delta\to0^+} \int \frac{\dd{\omega}}{2\pi} \frac{ \Sig_{pp}^\co\qty(\omega+\eps_p^\HF)}{(\omega + \ii \delta)^2} e^{-\ii \omega t}
\end{equation}
Substituting Eq.~\eqref{eq:Cpp} into the cumulant ansatz gives the diagonal elements of the Green's function in the time domain
\begin{equation} \label{eq:Gpp}
	G_{pp}(t) = - \ii \Theta(t) e^{-\ii \eps_p^\HF t + C_{pp}(t)} 
\end{equation}
The cumulant is often expressed as a function of the cumulant kernel
\begin{equation}
	\beta_p(\omega) = - \frac{1}{\pi} \Im\Sig^\co_{pp}\qty(\omega)
\end{equation}
as
\begin{equation}
	C_{pp}(t) 
	= 
	\int \dd{\omega} \frac{\beta_p\qty(\omega+\eps_p^\HF)}{\omega^2} \qty( e^{-\ii \omega t} + \ii \omega t - 1 )	
\end{equation}
This expression is known as the Landau form of the cumulant (see the \SupMat for a detailed derivation). \cite{Landau_1965}

\section{$GW$-based cumulant expansion}
\label{sec:GWC}
To derive the $GW$+C expression of the Green's function, let us now consider the diagonal elements of the $GW$ retarded self-energy, which reads
\begin{equation} \label{eq:SigC}
	\Sig_{pp}^\co(\omega) 
	= \sum_{i\nu} \frac{M_{pi\nu}^2}{\omega - \eps_i + \Om_\nu + \ii \eta} 
	+ \sum_{a\nu} \frac{M_{pa\nu}^2}{\omega - \eps_a - \Om_\nu + \ii \eta} 
\end{equation}
where the $\eps_p$'s are potentially quasiparticle energies depending on the level of self-consistency. 
The transition densities are given by
\begin{equation}
	\label{eq:sERI}
	M_{pq\nu} = \sum_{jb} \braket{pj}{qb} \qty(X_{jb,\nu} + Y_{jb,\nu})
\end{equation}
where $\braket{pq}{rs}$ are the usual two-electron integrals in Dirac notation and the matrices $\bX$ and $\bY$ gather the eigenvectors of the direct RPA problem \cite{SchuckBook,Chen_2017,Ren_2015} obtained via the following linear eigenvalue problem
\begin{equation}
\label{eq:RPA}
	\begin{pmatrix} 
    	\bA & \bB
		\\
    	-\bB & -\bA
	\end{pmatrix}
	\cdot
	\begin{pmatrix} 
    	\bX & \bY 
		\\
    	\bY & \bX 
    \end{pmatrix}
  	\\
	=
	\begin{pmatrix} 
    	\bX & \bY
		\\
    	\bY & \bX
    \end{pmatrix}
	\cdot
	\begin{pmatrix} 
    	\bOm & \bO
		\\
    	\bO & -\bOm
    \end{pmatrix}  
\end{equation}
where the diagonal matrix $\bOm$ contains the positive RPA eigenvalues $\Om_\nu$.
The matrix elements of the (anti)resonant block $\bA$ and the coupling block $\bB$ read 
\begin{subequations}
\begin{align}
	\label{eq:A}
    A_{ia,jb} & = (\eps_{a}-\eps_{i}) \delta_{ij}\delta_{ab} + \braket{ib}{aj} 
    \\
	\label{eq:B}
    B_{ia,jb} & =  \braket{ij}{ab}
\end{align}
\end{subequations}

Substituting the $GW$ retarded self-energy into Eq.~\eqref{eq:Cpp} and performing the frequency integration yields
\begin{equation} 
\begin{split} 
	C_{pp}(t) 
	& = \sum_{i\nu} \zeta_{pi\nu} \qty( e^{-\ii \Delta_{pi\nu} t} + \ii \Delta_{pi\nu} t - 1 )
	\\
	& + \sum_{a\nu} \zeta_{pa\nu} \qty( e^{-\ii \Delta_{pa\nu} t} + \ii \Delta_{pa\nu} t - 1 )
\end{split}
\end{equation}
with 
\begin{align}
	\zeta_{pi\nu} & = \qty(\frac{M_{pi\nu}}{\Delta_{pi\nu}})^2
	&
	\zeta_{pa\nu} & = \qty(\frac{M_{pa\nu}}{\Delta_{pa\nu}})^2
\end{align}
and
\begin{subequations}
\begin{align}
	\Delta_{pi\nu} & = \eps_i - \eps_p^\HF - \Om_\nu - \ii \eta
	\\
	\Delta_{pa\nu} & = \eps_a - \eps_p^\HF + \Om_\nu - \ii \eta
\end{align}
\end{subequations}

Therefore, defining 
\begin{equation}
	\eps_p^\QP = \eps_p^\HF + \Delta\eps_p^\QP
\end{equation}
where
\begin{equation}
	\Delta\eps_p^\QP 
	= - \sum_{i\nu} \Delta_{pi\nu} \zeta_{pi\nu} - \sum_{a\nu} \Delta_{pa\nu} \zeta_{pa\nu} 
	= \Sig_{pp}^\co\qty(\omega=\eps_p^\HF)
\end{equation}
is the quasiparticle shift, the diagonal elements of the Green's function in the frequency domain are given by
\begin{equation} \label{eq:Gpp_omega}
	G_{pp}(\omega) 
	= - \ii Z_p \int \dd{t} \Theta(t) e^{\ii \qty(\omega - \eps_p^\QP) t} 
		e^{\sum_{i\nu} \zeta_{pq\nu} e^{-\ii \Delta_{pi\nu} t} + \sum_{a\nu} \zeta_{pa\nu} e^{-\ii \Delta_{pa\nu} t}}
\end{equation}
where
\begin{equation} \label{eq:Zp_QP}
	Z_p^\QP 
	= \exp(- \sum_{i\nu} \zeta_{pi\nu} - \sum_{a\nu} \zeta_{pa\nu}) 
	= \exp(\eval{\pdv{\Sig_{pp}^\co(\omega)}{\omega}}_{\omega=\eps_p^\HF})
\end{equation}
is the quasiparticle weight (or renormalization factor) and the last term of Eq.~\eqref{eq:Gpp_omega} containing the double exponential is responsible for the appearance of satellites.
In the case of a $G_0W_0$ calculation where one linearizes the quasiparticle equation to obtain the quasiparticle energies, the $GW$+C renormalization factor associated with the quasiparticle peak and its $GW$ counterpart 
\begin{equation} \label{eq:Zp_GW}
	Z_p^\GW 
	= \frac{1}{1 - \eval{\pdv{\Sig_{pp}^\co(\omega)}{\omega}}_{\omega=\eps_p^\HF}}
\end{equation}
agree up to first order, as readily seen by comparing Eqs.~\eqref{eq:Zp_QP} and \eqref{eq:Zp_GW}.
Moreover, in this very specific case, it is easy to show that $\Re\qty( Z_p^\QP ) \le \Re\qty( Z_p^\GW )$, which evidences that the cumulant expansion systematically implies a redistribution of weights from the quasiparticle peak to the satellite structure. However, this is not always true in the general case.

Expanding the last term of Eq.~\eqref{eq:Gpp_omega} to first order, one obtains the following expression for the diagonal elements of the spectral function [see Eq.~\eqref{eq:Apq} for its definition]
\begin{equation} 
\begin{split} 
	A_{pp}(\omega) 
	\approx Z_p \delta\qty(\omega - \eps_p^\QP) 
	& + \sum_{i\nu} Z_{pi\nu}^\sat \delta\qty(\omega - \eps_{pi\nu}^\sat) 
	\\
	& + \sum_{a\nu} Z_{pa\nu}^\sat \delta\qty(\omega - \eps_{pa\nu}^\sat) + \cdots
\end{split}
\end{equation}
which features two sets of satellites at energies
\begin{align}
	\eps_{pi\nu}^\sat & = \eps_p^\QP + \Delta_{pi\nu} = \Delta \eps_p^\QP + \eps_i - \Om_\nu
	\\
	\eps_{pa\nu}^\sat & = \eps_p^\QP + \Delta_{pa\nu} = \Delta \eps_p^\QP + \eps_a + \Om_\nu
\end{align}
each located on a different branch and associated with the respective weights
\begin{align}
	Z_{pi\nu}^\sat & = Z_{p}^\QP \zeta_{pi\nu}
	&
	Z_{pa\nu}^\sat & = Z_{p}^\QP \zeta_{pa\nu}
\end{align}
where one can readily see that they are directly proportional to the quasiparticle spectral weight.
Here, we limit our analysis to these two sets of satellite peaks (especially the satellite peaks on the hole branch) as expanding to second order would produce satellites with even smaller weights and further away from the quasiparticle peak.

\begin{figure*}
	\includegraphics[width=0.8\linewidth]{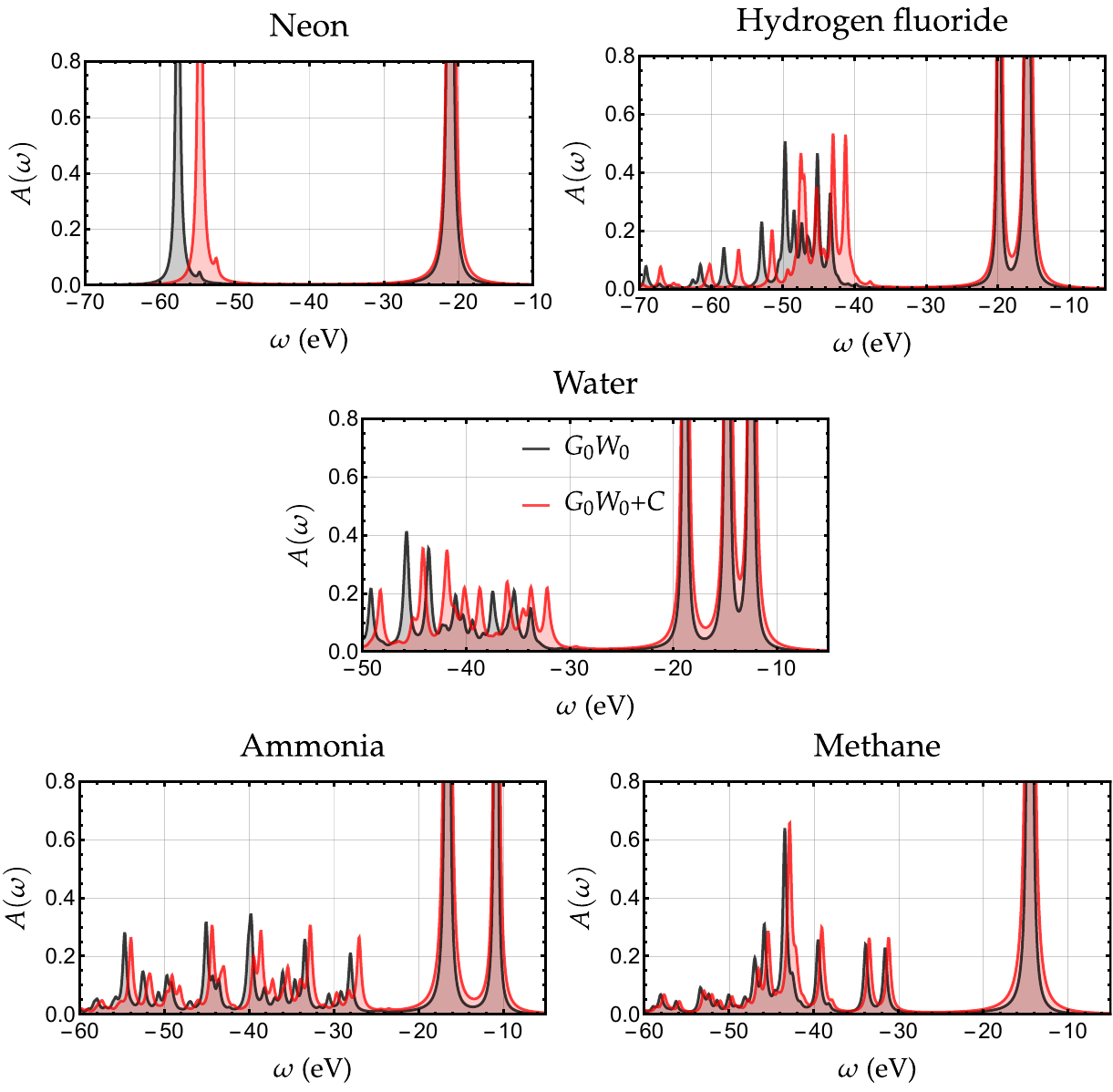}
\caption{
Outer-valence spectral functions of the 10-electron molecular series computed at the $G_0W_0$ (black) and $G_0W_0$+C (red) levels with the aug-cc-pVDZ basis and  $\eta = \SI{0.01}{\hartree}$}
\label{fig:A}
\end{figure*}

\begin{table*}
  \caption{Selection of quasiparticle energies of the 10-electron series computed at various levels of theory with the aug-cc-pVDZ basis and $\eta = \SI{0.001}{\hartree}$. The spectral weight is reported in parenthesis.}
  \label{tab:IPs}
  \begin{ruledtabular}
    \begin{tabular}{lllccccc}
      System & State   & Process        											&	$G_0W_0$		&	qs$GW$			&	$G_0W_0$+C		&	qs$GW$+C		&	FCI  \\
      \hline
		\ce{Ne}		& $1~^{2}\mathrm{P}$		&	$(2\mathrm{p})^{-1}$		&   21.104(0.947)	&	21.435(0.937)	&	20.983(0.942)	&	20.733(0.930)	&	21.426	\\
		\ce{HF}		& $1~^{2}\mathrm{\Sig}^+$	&	$(3\sig)^{-1}$				&   15.868(0.937)	&   16.144(0.924)	&   15.740(0.931)	&	15.510(0.916)	&	16.059	\\
					& $1~^{2}\mathrm{\Pi}$		&	$(2\mathrm{a}_1)^{-1}$		&	19.812(0.942)	&	20.084(0.931)	&	19.740(0.938)	&	19.497(0.926)	&	20.043	\\
		\ce{H2O}	& $1~^{2}\mathrm{B}_1$		&	$(1\mathrm{b}_1)^{-1}$		&   12.485(0.933)	&   12.640(0.920)	&   12.384(0.927)	&   12.228(0.912)	&   12.540	\\
					& $1~^{2}\mathrm{A}_1$		&	$(3\mathrm{a}_1)^{-1}$		&   14.781(0.935)	&   14.932(0.921)	&   14.698(0.929)	&   14.466(0.914)	&   14.829	\\
					& $1~^{2}\mathrm{B}_2$		&	$(1\mathrm{b}_2)^{-1}$		&   18.865(0.941)	&   19.069(0.931)	&   18.822(0.938)	&   18.706(0.928)	&   18.995	\\
		\ce{NH3}	& $1~^{2}\mathrm{A}_1$		&	$(3\mathrm{a}_1)^{-1}$		&   10.837(0.933)	&   10.870(0.922)	&   10.776(0.928)	&   10.663(0.915)	&   10.762	\\
					& $1~^{2}\mathrm{E}_1$		&	$(1\mathrm{e}_g)^{-1}$		&	16.578(0.940)	&	16.655(0.930)	&	16.544(0.936)	&	16.461(0.926)	&	16.534	\\
		\ce{CH4}	& $1~^{2}\mathrm{T}_2$		&	$(1\mathrm{t}_2)^{-1}$		&   14.466(0.943)	&   14.446(0.936)	&   14.445(0.940)	&	14.406(0.933)	&   14.285	\\
    \end{tabular}
  \end{ruledtabular}
\end{table*}

\begin{table*}
  \caption{Selection of satellite transition energies of the 10-electron series computed at various levels of theory with the aug-cc-pVDZ basis and $\eta = \SI{0.001}{\hartree}$. The IP-EOM-CCSDT and IP-EOM-CCSDTQ results from Ref.~\onlinecite{Marie_2024} are reported for comparison purposes.}
  \label{tab:Sat}
  \begin{ruledtabular}
    \begin{tabular}{lllcccccc}
			&		& 							& 			&		&		&	\mc{2}{c}{IP-EOM}			&	\\
																				\cline{7-8}
      System	& State	& Process						& $G_0W_0$	& $G_0W_0$+C	& qs$GW$+C	&	CCSDT	&	CCSDTQ	&	FCI	\\
      \hline
		\ce{Ne}		& $2~^{2}\mathrm{P}$ 		&	$(2\mathrm{p})^{-2}(3\mathrm{s})^1$							&	54.398	&	52.168	&	48.259	&			&			&	49.349	\\	
		\ce{HF}		& $2~^{2}\mathrm{\Delta}$	&	$(1\pi)^{-2}(4\sig)^1$										&	36.453	&	34.492	&	31.058	&	34.885	&	34.403	&	34.445	\\	
		\ce{H2O}	& $2~^{2}\mathrm{B}_1$		&   $(3\mathrm{a}_1)^{-1}(1\mathrm{b}_1)^{-1}(4\mathrm{a}_1)^1$	&   30.846  &	29.370  &	26.868	&   27.694	&	27.049	&	27.065	\\ 
					& $2~^{2}\mathrm{A}_1$		&   $(3\mathrm{a}_1)^{-2}(4\mathrm{a}_1)^1$						&   28.770  &	27.293  &   24.577	&   27.476	&	27.104	&	27.131	\\ 
					& $3~^{2}\mathrm{B}_1$		&	$(3\mathrm{a}_1)^{-1}(1\mathrm{b}_1)^{-1}(4\mathrm{a}_1)^1$	&   30.867  &	29.387	&   26.881	&   29.129	&	28.729	&	28.754	\\ 
		\ce{NH3}	& $2~^{2}\mathrm{A}_1$		&   $(3\mathrm{a}_1)^{-2}(4\mathrm{a}_1)^1$						&   24.410  &	23.510	&	21.657	&   24.101	&	23.818	&	23.829	\\ 
					& $2~^{2}\mathrm{E}$		&   $(3\mathrm{a}_1)^{-2}(3\mathrm{e})^1$						&   24.997  & 	24.098	&	22.317	&	25.882	&	25.648	&	25.655	\\ 
		\ce{CH4}	& $2~^{2}\mathrm{T}_2$		&	$(1\mathrm{t}_1)^{-2}(3\mathrm{a}_1)^1$						&	30.681	&	30.317	&	29.438	&	28.415	&	28.123	&	28.108	\\ 
    \end{tabular}
  \end{ruledtabular}
\end{table*}

\begin{figure*}
	\includegraphics[width=\linewidth]{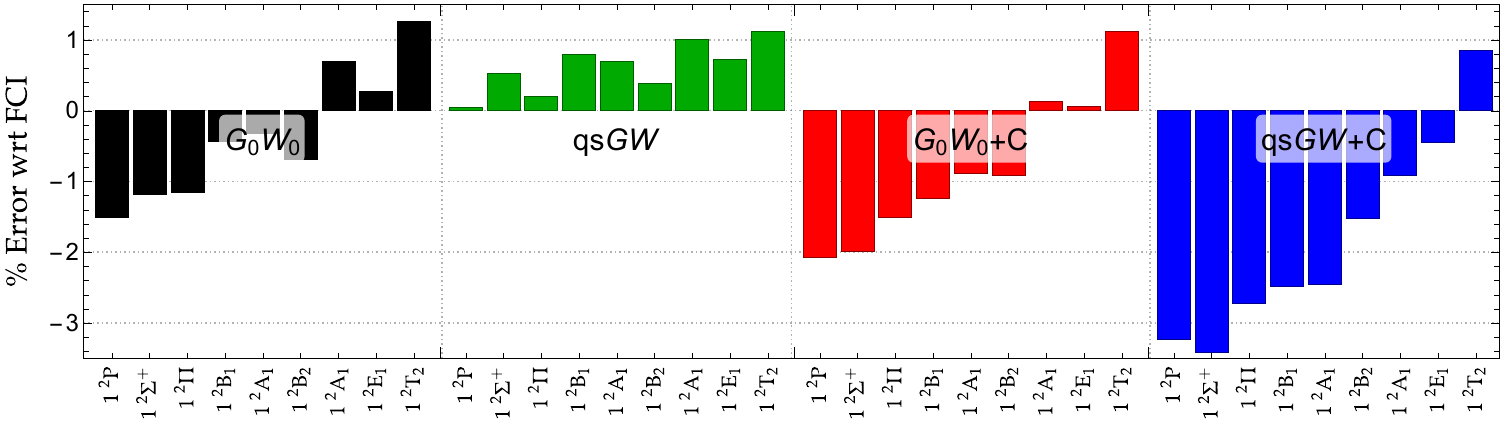}
\caption{
Percentage of error with respect to FCI for the outer-valence ionization energies reported in Table \ref{tab:IPs} computed at the $G_0W_0$, qs$GW$, $G_0W_0$+C, qs$GW$+C levels. All calculations are performed with the aug-cc-pVDZ basis and $\eta = \SI{0.001}{\hartree}$.}
\label{fig:IPs}
\end{figure*}

\begin{figure}
	\includegraphics[width=\linewidth]{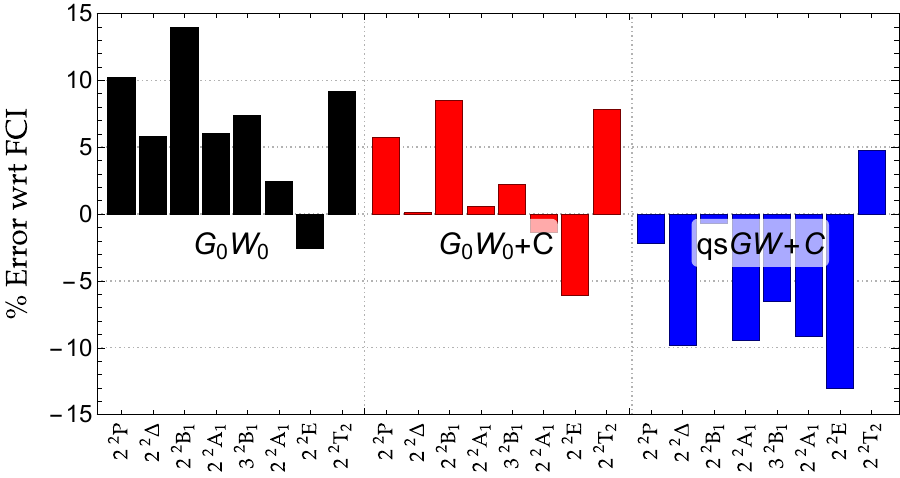}
\caption{
Percentage of error with respect to FCI for the satellite transition energies reported in Table \ref{tab:Sat} computed at the $G_0W_0$, $G_0W_0$+C, and qs$GW$+C levels. All calculations are performed with the aug-cc-pVDZ basis and $\eta = \SI{0.001}{\hartree}$}
\label{fig:Sat}
\end{figure}

\section{Spectral functions}
\label{sec:res}

The $GW$+C scheme has been implemented in \textsc{quack}, \cite{quack} an open-source electronic structure package for emerging methods.
The present implementation of $GW$ is described in Ref.~\onlinecite{Marie_2023b}.
To compute the quasiparticle energies at the one-shot $G_0W_0$ level, we do not linearize the self-energy, solving instead the frequency-dependent quasiparticle equation using Newton's method:
\begin{equation} \label{eq:qp_eq}
	\omega - \eps_p^\HF - \Re\qty[\Sig_{pp}^{\co}(\omega)] = 0
\end{equation}
In the $G_0W_0$ calculations, we set $\eta = \SI{0.001}{\hartree}$ in the expression of the self-energy [see Eq.~\eqref{eq:SigC}] unless we plot the spectral function in which case we set $\eta = \SI{0.01}{\hartree}$ to broaden the peaks further (see below). Note also that, at the $G_0W_0$ level, we have $\eps_p = \eps_p^\HF$ in the various expressions reported in Sec.~\ref{sec:cumulant}.
The HF starting point quantities are computed in the restricted formalism.
The qs$GW$ calculations are performed with the regularized scheme based on the similarity renormalization group approach, as described in Ref.~\onlinecite{Marie_2023}. A flow parameter of $s = 500$ is employed. In the case of self-consistent calculations, we have $\eps_p = \eps_p^\GW$.
All (hole and particle) states are corrected.
We systematically employed Dunning's aug-cc-pVDZ basis set for all calculations and consider below the well-known 10-electron series of molecular systems, namely, \ce{Ne}, \ce{HF}, \ce{H2O}, \ce{NH3}, and \ce{CH4}.
As reference for the outer-valence quasiparticle and satellite energies, we rely on the full configuration interaction (FCI) values reported in Ref.~\onlinecite{Marie_2024}, from where the geometries have also been extracted.

At the $G_0W_0$ level, satellite transition energies have been obtained via the linear version of the $GW$ equations as described in Refs.~\onlinecite{Bintrim_2021,Monino_2022,Monino_2023,Scott_2023}. Within this scheme,  the so-called ``upfolded'' matrix is diagonalized in a larger space that includes the 2h1p and 2p1h configurations, the satellite energies being obtained as higher/lower roots.  
Although this procedure is more computationally expensive than the usual ``downfolded'' version where one solves the non-linear equation \eqref{eq:qp_eq}, it is still technically feasible for the present systems and ease the obtention of satellite energies which are extremely challenging to get as solutions of the non-linear equation. Satellite energies are not computed at the qs$GW$ level due to the static nature of this approximation.

At the $GW$ level, the spectral function is
\begin{equation}
	A_p^\GW(\omega) = - \frac{1}{\pi} \frac{\Im\qty[\Sig_{pp}^\co(\omega)]}{ \qty{ \omega - \eps_p^\HF - \Re\qty[\Sig_{pp}^\co(\omega)] }^2 + \qty{\Im\qty[\Sig_{pp}^\co(\omega)]}^2}
\end{equation}
where the expression of the elements $\Sig_{pp}^\co(\omega)$ is given in Eq.~\eqref{eq:SigC}. 
At the $GW$+C level, the spectral function reads (see \SupMat)
\begin{equation}
	A_p^\GWC(\omega) = A_p^\QP(\omega) + \sum_{i\nu} A_{pi\nu}^\C(\omega) + \sum_{a\nu} A_{pa\nu}^\C(\omega) + \ldots
\end{equation}
where the quasiparticle part is
\begin{equation}
	A_{p}^\QP(\omega) = - \frac{1}{\pi} \frac{\Re\qty(Z_p^\QP) \Im\qty(\eps_p^\QP) + \Im\qty(Z_p^\QP) \qty[ \omega - \Re\qty(\eps_p^\QP) ] }{ \qty[ \omega - \Re\qty(\eps_p^\QP) ]^2 + \qty[\Im\qty(\eps_p^\QP)]^2}
\end{equation}
and the satellite contributions are given by the following expressions:
\begin{subequations}
\begin{align}
	A_{pi\nu}^\C(\omega) & = - \frac{1}{\pi} \frac{\Re\qty(Z_{pi\nu}^\sat) \Im\qty(\eps_{pi\nu}^\sat) + \Im\qty(Z_{pi\nu}^\sat) \qty[ \omega - \Re\qty(\eps_{pi\nu}^\sat) ] }{ \qty[ \omega - \Re\qty(\eps_{pi\nu}^\sat) ]^2 + \qty[\Im\qty(\eps_{pi\nu}^\sat)]^2}
	\\
	A_{pa\nu}^\C(\omega) & = - \frac{1}{\pi} \frac{\Re\qty(Z_{pa\nu}^\sat) \Im\qty(\eps_{pa\nu}^\sat) + \Im\qty(Z_{pa\nu}^\sat) \qty[ \omega - \Re\qty(\eps_{pa\nu}^\sat) ] }{ \qty[ \omega - \Re\qty(\eps_{pa\nu}^\sat) ]^2 + \qty[\Im\qty(\eps_{pa\nu}^\sat)]^2}
\end{align}
\end{subequations}

The outer-valence spectral function $A(\omega) = \sum_{p} A_p(\omega)$ (where the sum is performed on the outer-valence orbitals) computed at the $G_0W_0$ and $G_0W_0$+C levels for each 10-electron molecular system is represented in Fig.~\ref{fig:A}.
For example, the spectral function of water features three well-defined quasiparticle peaks around 13, 15, and \SI{19}{\eV} and satellite structures higher in energy.
Figure \ref{fig:A} evidences that the main difference between the $G_0W_0$ and $G_0W_0$+C spectral functions is a global shift to lower energy of the satellite features, the quasiparticle peaks being much less affected. In addition, the intensities of the satellite peaks are also slightly affected but the overall structure of the satellite band is not modified by the cumulant expansion.
On the other hand, vertex corrections to the $G_0W_0$ scheme have been shown to create new satellite features in ionization spectra (see Ref.~\onlinecite{Mejuto-Zaera_2021}).

Tables \ref{tab:IPs} and \ref{tab:Sat} report the outer-valence ionization energies and satellite transition energies, respectively, for the present set of molecules. 
Most of these satellites have vanishing weights.
The percentage of error with respect to the reference FCI values is represented in Figs.~\ref{fig:IPs} and \ref{fig:Sat}.
The first ionization of water corresponds to an electron removal from the orbital $1b_1$, and this process is denoted as $(1\mathrm{b}_1)^{-1}$.
The lowest satellite transition in \ce{H2O} corresponds to the process $(3\mathrm{a}_1)^{-1}(1\mathrm{b}_1)^{-1}(4\mathrm{a}_1)^1$, where one electron is removed from orbitals $3\mathrm{a}_1$ and $1\mathrm{b}_1$ and one is attached in orbital $4\mathrm{a}_1$. The symmetry labels of these charged excited states are also reported in Tables \ref{tab:IPs} and \ref{tab:Sat}.

Figure \ref{fig:IPs} shows that the quasiparticle energies are not improved by the cumulant expansion. They are actually slightly worse. It is particularly true at the qs$GW$ level where the error is increased by 1 or 2\% when one considers the cumulant scheme. 
The spectral weights of the quasiparticle peaks reported in Table \ref{tab:IPs} show that the cumulant induces a small redistribution of spectral weight from the quasiparticle to the satellites but this effect is quite subtle.

Concerning the satellites (see Fig.~\ref{fig:Sat}), we find that $G_0W_0$+C can significantly reduce the error of $G_0W_0$ in certain situations. This is the case for the $2~^{2}\mathrm{\Delta}$ satellite of \ce{HF} (from \SI{2}{\eV} to \SI{0.05}{\eV}), the $1~^{2}\mathrm{A}_1$ and $3~^{2}\mathrm{B}_1$ satellites of \ce{H2O} (from \SI{1.6}{\eV} to \SI{0.2}{\eV} and from \SI{2}{\eV} to \SI{0.6}{\eV}), and to a lesser extent, the $2~^{2}\mathrm{A}_1$ satellite of \ce{NH3} (from \SI{0.6}{\eV} to \SI{-0.3}{\eV}). On the contrary, the qs$GW$+C scheme tends to overcorrect the satellite energies.
The $2~^{2}\mathrm{P}$ satellite of \ce{Ne}, the $2~^{2}\mathrm{B}_1$ satellite of \ce{H2O}, the $2~^{2}\mathrm{E}$ satellite of \ce{NH3}, and the $2~^{2}\mathrm{T}_2$ satellite of \ce{CH4} are much harder to describe at the $GW$ level and errors remain larger even within the cumulant scheme.

\section{Concluding Remarks}
\label{sec:ccl}
In the present article, we have reviewed the literature on cumulant Green's function methods, provided a detailed derivation of the associated equations, and investigated the performance of this scheme in the context of molecular systems where solid reference data for satellite transitions are now available.
In particular, we have compared the satellite transition energies obtained via the $GW$+C scheme and the ``upfolded'' version of the $GW$ equations. 
In a nutshell, $G_0W_0$+C does sometimes improve upon $G_0W_0$ but it is far from being systematic. The cumulant version of qs$GW$ has been found to usually overcorrect satellite energies.
However, the cumulant approach allows us to estimate satellite energies without solving the dynamical $GW$ equations or its larger (hence more expensive) upfolded version. 
These observations show that the development of new Green's function methods capable of describing accurately satellite states in molecular systems would be extremely useful.

\section*{Conflicts of interest}
There are no conflicts to declare.

\section*{Acknowledgements}
This project has received funding from the European Research Council (ERC) under the European Union's Horizon 2020 research and innovation programme (Grant agreement No.~863481). Additionally, it was supported by the European Centre of Excellence in Exascale Computing (TREX), and has received funding from the European Union's Horizon 2020 --- Research and Innovation program --- under grant agreement no.~952165.

\section*{References}

\bibliography{cumulant}

\end{document}